\documentclass[traditabstract, final]{aa}
\usepackage{graphicx}
\usepackage[small]{caption}
\usepackage{epsfig}
\usepackage{natbib}
\usepackage{fancyhdr}
\usepackage{amsmath}
\usepackage{amssymb}
\usepackage{makeidx}
\usepackage{titletoc}
\usepackage[]{subfigure}
\usepackage{lscape}
\usepackage{hyperref}
\usepackage{wasysym}
\usepackage{wrapfig}
\usepackage{longtable}
\usepackage{aas_macros}
\usepackage{txfonts}

\begin{document}

\title{A new study of stellar substructures in the Fornax dwarf spheroidal galaxy}
\titlerunning{A new study of stellar substructures in the Fornax dwarf spheroidal galaxy}

   \author{T.J.L. de Boer\inst{1}
          \and
          E. Tolstoy\inst{1} 
          \and
          A. Saha\inst{2}
          \and
          E.W. Olszewski\inst{3}
          }

   \institute{Kapteyn Astronomical Institute, University of Groningen,
              Postbus 800, 9700 AV Groningen, The Netherlands\\
              \email{deboer@astro.rug.nl}
             \and
              National Optical Astronomy Observatory\thanks{The National Optical Astronomy Observatory is operated by AURA, Inc., under cooperative agreement with the National Science Foundation.},
              P.O. box 26732, Tucson, AZ 85726, USA
             \and
             Steward Observatory, The University of Arizona, Tucson, AZ 85721, USA
             }

   \date{Received ...; accepted ...}

\abstract{Using deep V, B$-$V wide-field photometry, we have conducted a new study of stellar over-densities in the Fornax dwarf spheroidal galaxy by determining detailed Star Formation Histories from Colour-Magnitude Diagram analysis. We have concentrated on the relatively young stellar component ($<$4 Gyr old), and compared this to the underlying Fornax field population. \\
We have studied in more detail the previously known inner shell-like structure and shown that it has a well-defined age-metallicity relation with a peak at $\approx$1.5 Gyr, [Fe/H]=$-$0.6 dex. Comparison to the Fornax centre shows that the over-dense feature is consistent with the age-metallicity relation of young field stars, and likely formed from Fornax gas. This is consistent with a scenario in which the over-density was formed by the re-accretion of previously expelled gas. \\
We have also discovered a new stellar over-density, located 0.3 degrees~(0.7 kpc) from the centre, which is only 100 Myr old, with solar metallicity. This feature constitutes some of the youngest, most metal-rich stars stars observed in Fornax to date. It is unclear how the young over-density was formed, although the age and metallicity of stars suggest this feature may represent the last star formation activity of the Fornax dSph. 
}


\maketitle

\section{Introduction}
\label{Fnxintroduction}
The Fornax dwarf spheroidal galaxy~(dSph) has experienced an unusually complex star formation history~(SFH), with stellar populations covering a large range in age and metallicity~\citep[e.g.,][]{Gallart052, Coleman08, deBoer2012B}. Spectroscopic studies of Red Giant Branch~(RGB) stars have found the Metallicity Distribution Function~(MDF) to have a broad distribution with a dominant metal-rich~([Fe/H]$\approx$$-$0.9 dex) component~\citep[e.g.,][]{Pont04, Battaglia06, Letarte10}. The kinematics of these RGB stars have shown the presence of at least three kinematically distinct populations~\citep{Battaglia06, Amorisco12}. Analysis of the stellar spatial distribution shows that Fornax contains a radial population gradient, in which younger, more metal-rich stars are found progressively more towards the centre~\citep{Battaglia06, Coleman08, deBoer2012B}. \\
Wide-field photometric surveys of Fornax have previously found two young stellar over-densities, which have been interpreted as shell-like features resulting from the in-fall of a smaller system less than $\sim$2 Gyr ago~\citep{Coleman04,Coleman052,Olszewski06}. The inner feature is located at a distance of $\approx$17 arcmin~(0.7 kpc) from the centre of Fornax, while the outer feature is located outside the tidal radius at a distance of 1.3 degrees~(3 kpc) from the centre. Isochrone fitting to the observed Colour Magnitude Diagram~(CMD) determined that the main population of both over-densities formed less than 2 Gyr ago~\citep{Coleman052,Olszewski06}. Metallicities determined from the CMD suggest that the stellar populations of the inner feature are similar to the young stars of the Fornax field stars, indicating a connection between both components~\citep{Olszewski06}. \\
It is still unclear what has driven the complex evolution history of Fornax. Models predicting the formation of Fornax have tried to explain the spatial complexity and extended, multiple episodes of star formation using tidal encounters with the Milky Way, or re-accretion of expelled gas~\citep[e.g.,][]{Pasetto11, Nichols12, Yozin12}. Models involving accretion events are especially relevant, given the presence of these shell-like structures in and around Fornax. \\
If the over-dense features were the result of an in-fall event, it is likely that the interaction had an effect on the populations of the over-density and the Fornax field. Therefore, the detailed properties of the stellar populations in the over-densities can be used to constrain the nature of the in-falling system. If the in-falling system was a small gas-rich galaxy, the over-densities will contain a mix of populations made up of stars already present in the accreted galaxy and new populations produced as a result of the interaction. Since smaller galaxies are usually metal-poor, and young stellar populations are intrinsically blue, the over-density should contain populations different from the host galaxy onto which it is accreted~\citep{Pence86,McGaugh90}. \\
If the over-dense features were created by the accretion of a gas cloud onto the host galaxy, the stellar population properties will depend on the metallicity of the in-falling gas. The in-fall of a cloud of pristine gas would result in an over-density containing metal-poor, blue populations while the re-accretion of previously enriched gas would lead to redder populations. Finally, if the features are part of the standard Fornax field population, the stellar populations would match the Age-Metallicity Relation~(AMR) of the Fornax field stars~\citep{deBoer2012B}.\\
In this paper we use the photometric dataset presented in~\citet{deBoer2012B} to study the known over-density and search for new features in the spatial distribution of young stars in Fornax. The deep CMDs of the stellar over-densities were analysed using the technique described in~\citet{deBoer2012A}, to determine the ages and metallicities of the over-dense populations and obtain their AMR. By comparing the AMR of the over-densities to the Fornax field, we can constrain the likely nature and origin of the over-dense features and their role in the formation of Fornax. \\
The paper is structured as follows: in Section~\ref{targetselection} we analyse the photometric dataset to search for photometric over-densities in the young stars of Fornax. In Section~\ref{shellCMD} we present the CMDs for the over-densities and their associated field component. Section~\ref{FnxSFH} presents the SFH results for the over-densities, which are compared to the Fornax centre in Section~\ref{centrecomparison}. Finally, Section~\ref{conclusions} discusses the possible formation scenarios for the over-densities and their effect on the formation and evolution of the Fornax dSph.
\begin{figure}[!ht]
\centering
\includegraphics[angle=0, width=0.49\textwidth]{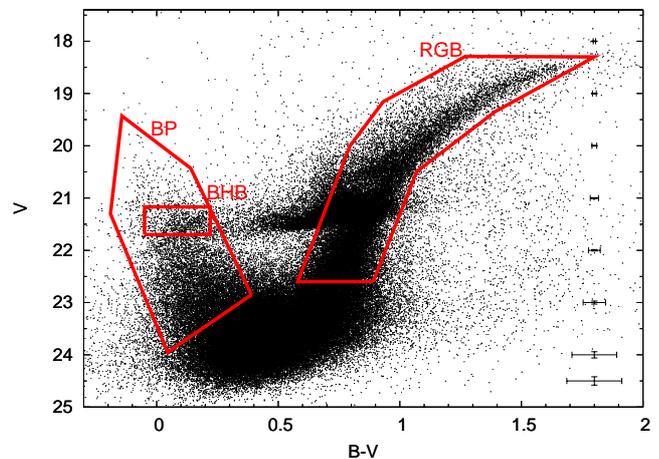}
\caption{(V, B$-$V) CMD of the Fornax centre, indicating the regions corresponding to the old RGB field population, young BP stars and old BHB stars. \label{CMDselection}} 
\end{figure}
\begin{figure*}[!ht]
\centering
\includegraphics[angle=0, width=0.49\textwidth]{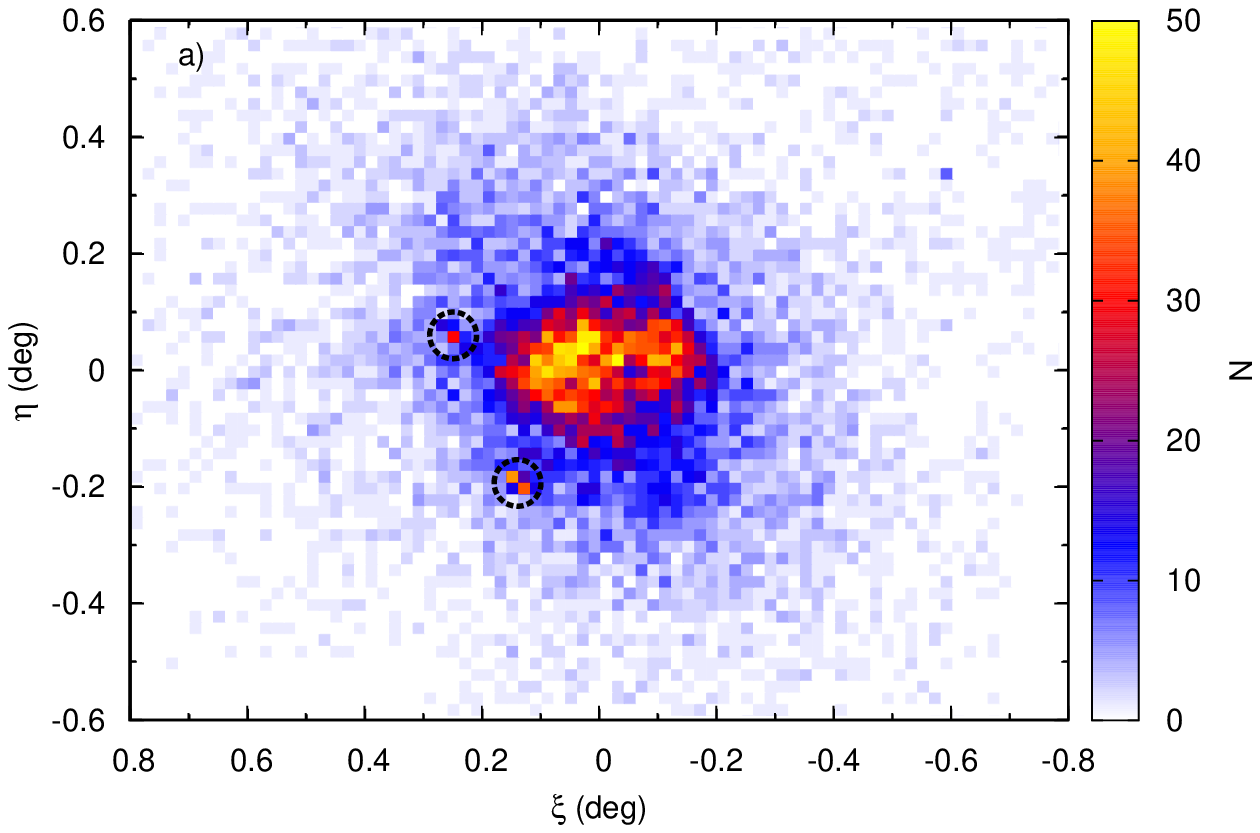}
\includegraphics[angle=0, width=0.49\textwidth]{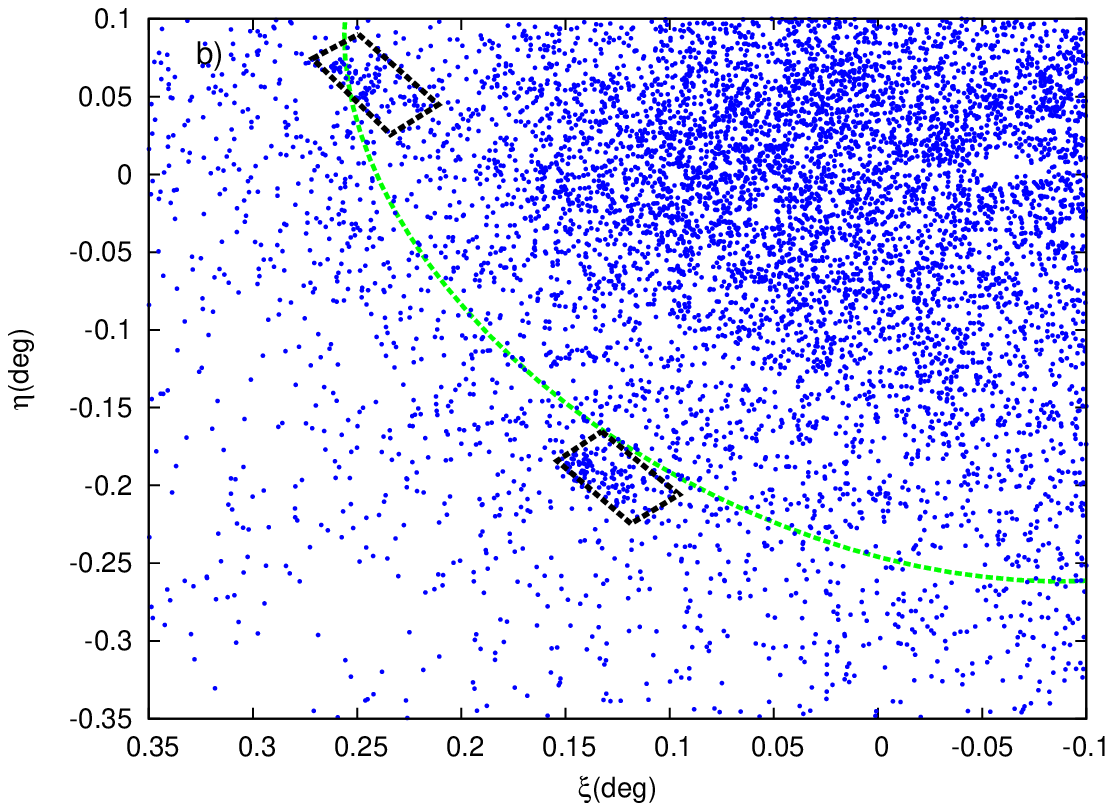}
\caption{\textbf{a)} The spatial Hess$-$like diagram of BP stars in the Fornax dSph. The two dashed circles indicate the position of two over-densities, one known and one new. \textbf{b)} The spatial distribution of BP stars in the region corresponding to both detected over-densities. Dashed~(black) boxes indicate the area selected for each over-density. The~(green) ellipse shows the annulus corresponding to a radius of r$_{ell}$=0.3 degrees from the centre of Fornax, assuming the structural parameters from~\citet{Irwin95}. \label{spatialBPhess}} 
\end{figure*}

\vspace{-0.25cm}
\section{Over-density selection}
\label{targetselection}
We used the wide-field photometric survey of the Fornax dSph described in detail in~\citet{deBoer2012B} to investigate the presence of young stellar over-densities in Fornax. This survey consisted of deep optical photometry obtained using the CTIO 4-m MOSAIC II camera, covering a large part of the galaxy in the B,V and I filters. The photometric catalog is complete out to elliptical radius r$_{ell}$$\le$0.8 degrees for the B and V filters, while the I filter is complete only for r$_{ell}$$\le$0.4 degrees. The elliptical radius is defined as the major axis length of an ellipse centred on Fornax with structural parameters according to~\citet{Irwin95}. \\
To select young over-densities in Fornax, we investigate the spatial distribution of young Main Sequence stars on the Blue Plume~(BP), which stand out against the predominantly older RGB field population~(see Figure~\ref{CMDselection}). These stars are selected from the CMD, with the area corresponding to the blue horizontal branch removed to avoid contaminating the young sample with old, metal-poor stars. Figure~\ref{spatialBPhess}a shows the resulting spatial Hess$-$like diagram of young MS stars in Fornax, obtained by binning the spatial coordinates of each star. The young population is distributed in a nearly horizontal structure with two main over-dense regions, one of which is located very close to the position of Fnx~4~\citep{Mackey03}. This is clearly different from the intermediate and old populations, which are distributed in an ellipse with position angle of 46.8 degrees~\citep{Stetson98, Saviane00, Battaglia06}. \\
The over-density discovered by~\citet{Coleman04} is clearly visible in Figure~\ref{spatialBPhess}a, at $\xi$$\approx$0.13 deg, $\eta$$\approx$$-$0.2 deg. The position and alignment of the feature is consistent with the parameters given by~\citet{Coleman04}. The current spatial coverage of our deep photometry is not enough to be able to see the outermost feature, located outside the tidal radius of Fornax at a distance of 1.3 degrees from the centre~\citep{Coleman05}. However, another over-dense feature stands out clearly in the spatial distribution of young stars, East of the Fornax centre, at $\xi$=0.24 deg, $\eta$=0.06 deg (RA=02$^{h}$:41$^{m}$:02.2$^{s}$, DEC=-34$^{\circ}$:27$^{\prime}$:17.6$^{\prime\prime}$). This feature is located at a distance of 0.3 degrees, or 0.7 kpc from the centre of Fornax. \\
Figure~\ref{spatialBPhess}b shows a zoom in of the distribution of young stars in the inner region of Fornax, where both over-densities are clearly visible. The sample of stars adopted for the known over-density is selected according to the structural properties derived by~\citet{Coleman04}. For the new over-density we adopt a rectangular region with dimensions of 1.7$^{\prime}$$\times$3.6$^{\prime}$ at a position angle of PA=36 deg. \\
The reason why the new over-density was not found in previous Fornax surveys is mainly due to the photometric depth and available filters. Most surveys~\citep[e.g.,][]{Eskridge883, Eskridge884, Irwin95, Walcher03, Coleman052} did not extend deep enough or were not able to make a colour selection to obtain a clean sample of young stars, which dominate the young over-density. The survey by~\citet{Olszewski06} did go deep enough, but did not cover the region of the new over-density.
\begin{figure*}[!ht]
\centering
\includegraphics[angle=0, width=0.49\textwidth]{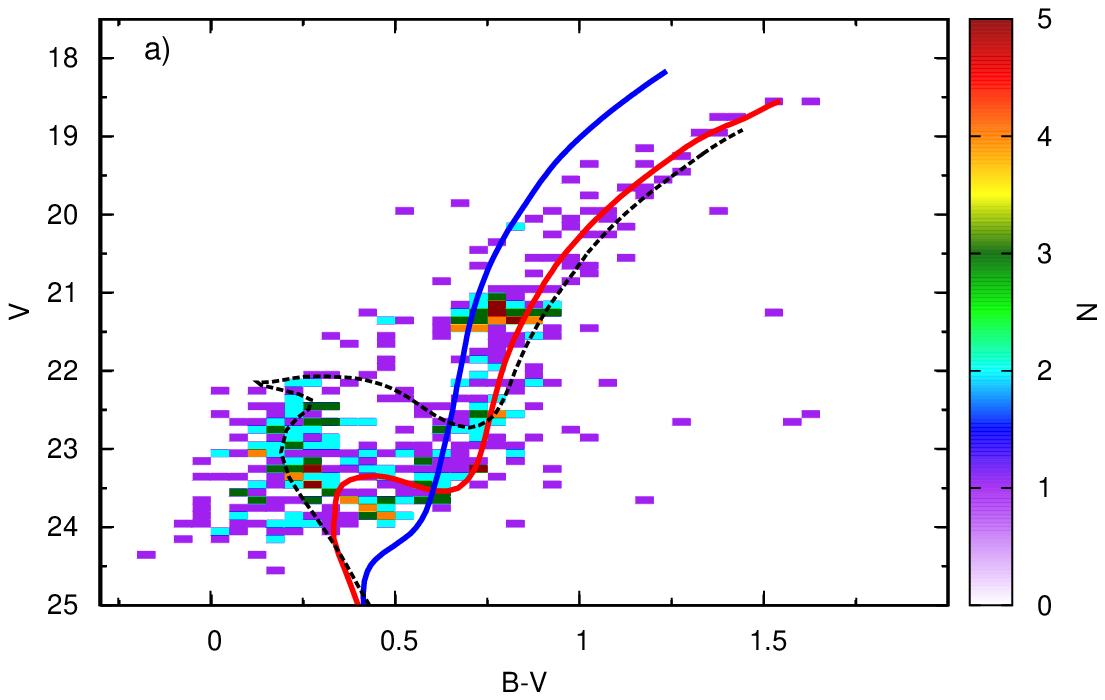}
\includegraphics[angle=0, width=0.49\textwidth]{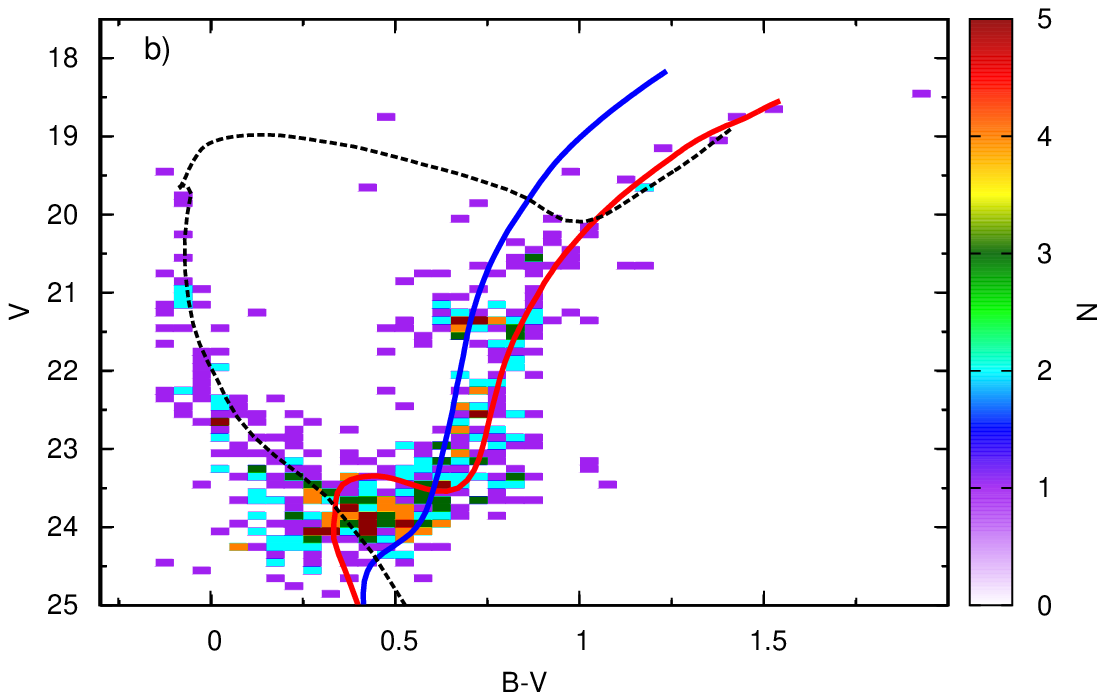}
\caption{The Hess diagram of~\textbf{a)} the known over-density found by~\citet{Coleman04} and~\textbf{b)} the new over-density. Three isochrones are overlaid, representative of the populations traced by the Fornax field RGB~(blue and red), as well as the over-dense population in the both features~(black, dashed). The range of older field populations is traced by a metal-rich~([Fe/H]=$-$1.00 dex, [$\alpha$/Fe]=0.00 dex, age=4 Gyr, red) and metal-poor~([Fe/H]=$-$2.45 dex, [$\alpha$/Fe]=0.40 dex, age=14 Gyr, blue) isochrone. Parameters of the overlaid isochrone of the over-dense population are [Fe/H]=$-$0.60 dex, [$\alpha$/Fe]=0.00 dex, age=1.5 Gyr in panel~\textbf{a)} and [Fe/H]=0.00 dex, [$\alpha$/Fe]=0.00 dex, age=0.2 Gyr in panel~\textbf{b)}. \label{shells}} 
\end{figure*}
\begin{figure*}[!ht]
\centering
\includegraphics[angle=0, width=0.49\textwidth]{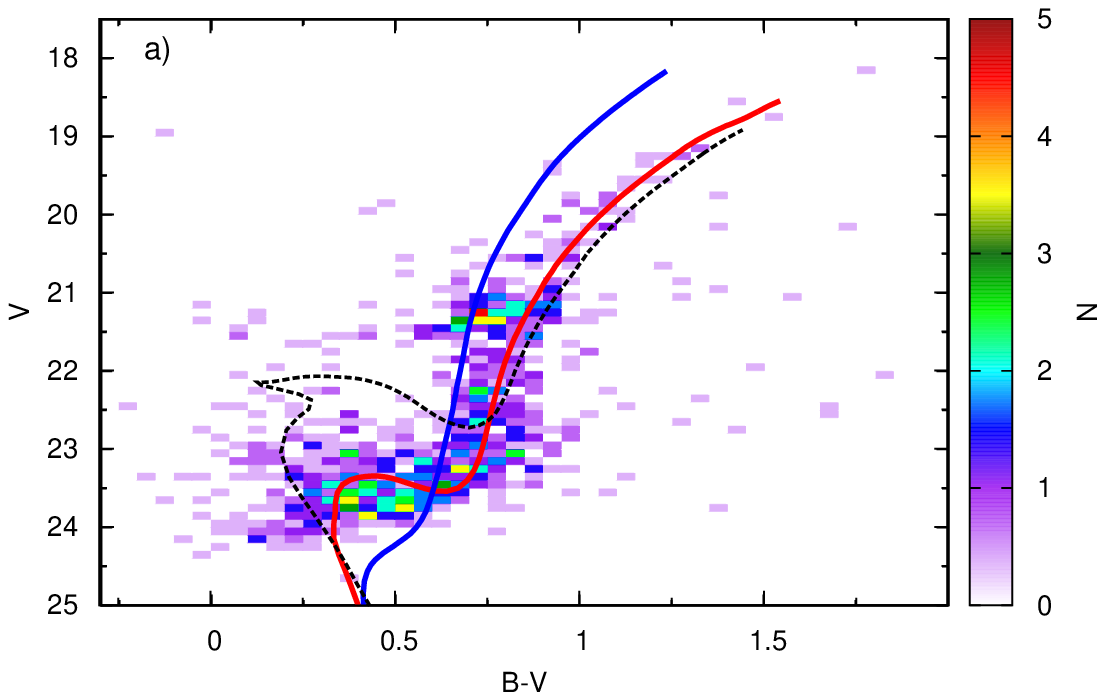}
\includegraphics[angle=0, width=0.49\textwidth]{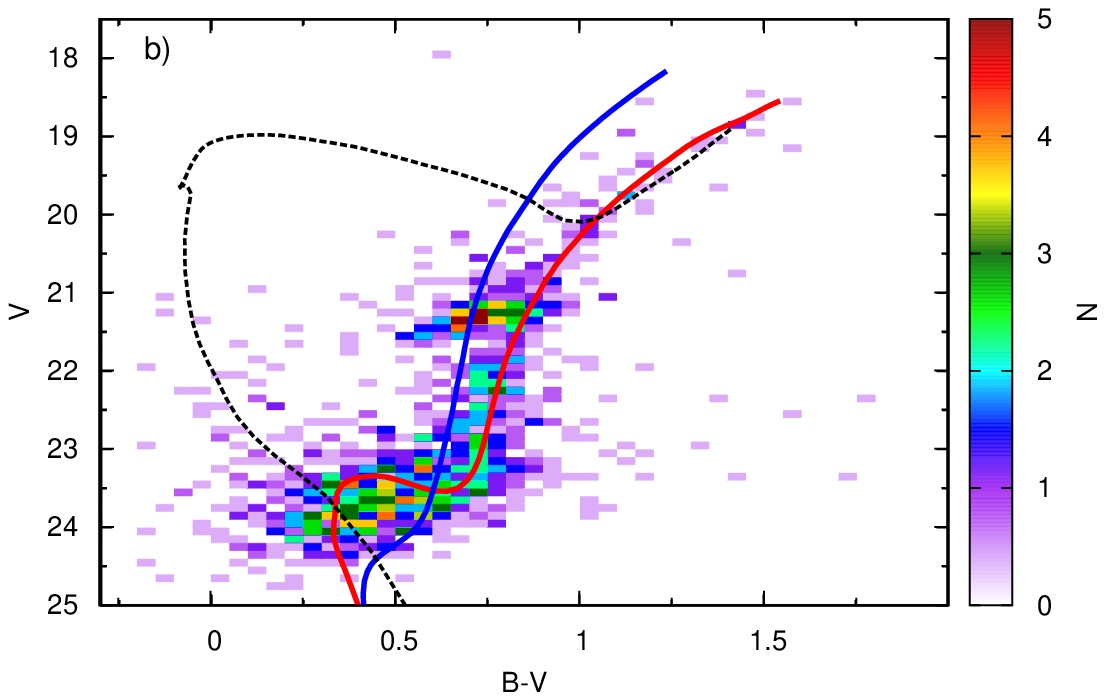}
\caption{Hess diagrams of the Fornax field star contribution associated to~\textbf{a)} the known over-density and~\textbf{b)} the new over-density. The same three isochrones as described in Figure~\ref{shells} are overlaid on the Hess diagram, representative of the main populations traced by the average Fornax field, and over-dense populations . \label{shellsBG}} 
\end{figure*}

\vspace{-0.25cm}
\subsection{Fornax field star contamination}
\label{BGselection}
Due to the position of the over-densities within Fornax, it is likely that the CMD of each feature is contaminated by stars belonging to the Fornax field. To determine an accurate SFH for the over-density, it is necessary to account for this contamination. Given the fact that the Fornax dSph displays different stellar populations at different elliptical radii~\citep{deBoer2012B}, we determine the field star contribution by only taking into account stars with the same range of elliptical radii as the studied region. Within this annulus, we select only stars located far from the over-density~(more than twice the extent). The selected region is expanded until it contains a sufficient number of stars to properly sample the CMD~(typically 1-2 times the number of stars in the over-density). Finally, the number of stars is scaled to the same spatial area as the studied over-density, to construct a CMD representative of the Fornax field population. 
\begin{figure*}[!ht]
\centering
\includegraphics[angle=0, width=0.97\textwidth]{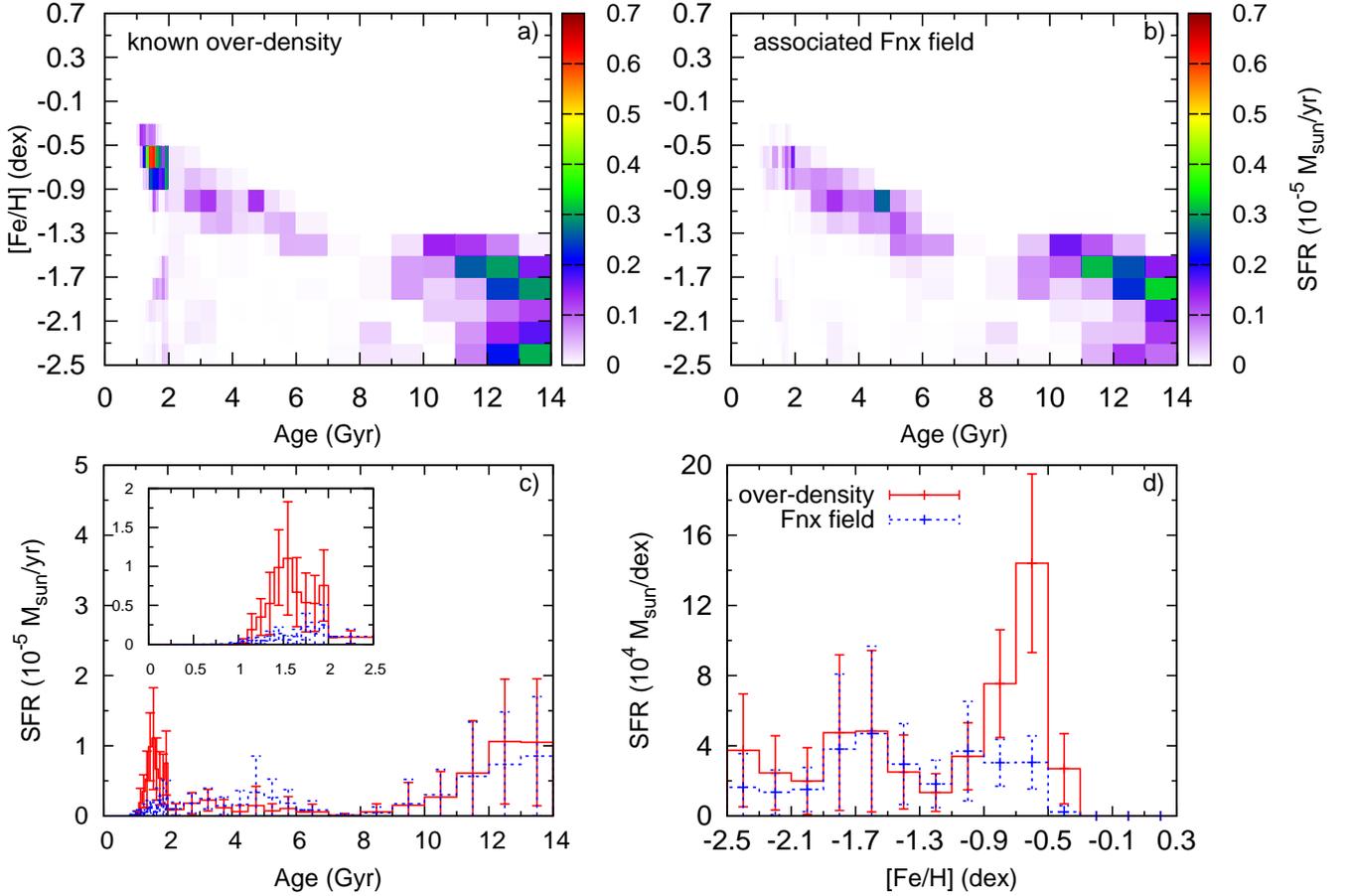}
\caption{The SFH of the known over-density, as obtained from the V,B-V CMD. The upper row shows the full 2D solution for the over-density~(\textbf{a}) and the associated Fornax field star contribution~(\textbf{b}), for comparison. The bottom row shows the SFH~(\textbf{c}) and CEH~(\textbf{d}) derived of the over-density~(red, solid histogram) as well as the Fornax field~(blue, dashed histogram). A zoom-in of the SFH at the youngest ages is also shown. \label{FnxSFHshell}} 
\end{figure*}

\section{Colour-magnitude diagrams}
\label{shellCMD}
Using the regions selected in Section~\ref{targetselection}, we construct the CMD of each over-density and compare it to the CMD of the associated Fornax field star contribution. Figure~\ref{shells} shows the observed CMD for the two over-densities in the form of Hess diagrams~(a CMD density plot). The Hess diagrams are overlaid with isochrones from the Yonsei$-$Yale isochrone set~\citep{YonseiYaleI}, representative of intermediate and old Fornax field RGB populations~([Fe/H]=$-$1.00 dex, [$\alpha$/Fe]=0.00 dex, age=4.0 Gyr and [Fe/H]=$-$2.50 dex, [$\alpha$/Fe]=0.40 dex, age=14.0 Gyr respectively). A reference isochrone indicating the position of the main over-dense population is also shown. The observed Hess diagram consists of a combination of stars from the over-dense populations and from the Fornax field. The Fornax field star contribution associated to each over-density is shown in Figure~\ref{shellsBG}, scaled to the same spatial area as the observed sample. \\
The CMD in Figure~\ref{shells}a shows several features similar to those of the Fornax field, such as a broad, metal-rich RGB~(traced by the red isochrone), a thin red horizontal branch and prominent red clump. Figure~\ref{shells}a also shows a prominent young MS population~(black, dashed line), with a broad, well developed turn-off around V$\approx$22.5. This feature is only weakly visible in the Fornax field~(Figure~\ref{shellsBG}), while the number of intermediate and old RGB stars in both CMDs is comparable. Comparison to the reference isochrone~(black, dashed line) in Figure~\ref{shells}a shows that the young population is consistent with an age of $\approx$1.5 Gyr~\citep{Olszewski06}. \\
The CMD of the new over-density~(Figure~\ref{shells}b) is also dominated by stars from the Fornax field, as shown in Figure~\ref{shellsBG}b. However, a very young, blue Main Sequence~(MS) population is also visible, blueward of the main Fornax MSTO at B$-$V=0.2. This feature is not reproduced in the field star contribution, and is responsible for the clear over-density in Figure~\ref{spatialBPhess}a. The young MS extends from the Fornax field MSTO at V=23.5 up to V=19.5, containing 80 stars with V$\le$23.5, B$-$V$\le$0.2~(compared to 18 in the associated Fornax field). Comparison to stellar evolution models shows that this young MS is consistent with an age of 0.2 Gyr~(black, dashed line). This is substantially younger than the populations found in the known over-densities, and constitutes some of the youngest star formation in the Fornax dSph. 

\section{Star formation history}
\label{FnxSFH}
The SFH of the young over-densities is determined using the SFH fitting method Talos, described in detail in~\citet{deBoer2012A}. For both over-densities, the distance modulus~[(m-M)$_{\mathrm{V}}$=20.84] and extinction value~(E(B-V)=0.03) is assumed to be the same as for the Fornax field stars~\citep{Schlegel98, Pietrzynski09}. Given the young ages expected for stars in the substructures, we adopt the Yonsei$-$Yale isochrone set~\citep{YonseiYaleI} in the SFH fitting, since it includes isochrones with ages going down to 1 Myr and a broad metallicity range. For the SFH solution, metallicities are allowed to range from $-$2.5$\le$[Fe/H]$\le$$+$0.3 dex with a spacing of 0.2 dex, for ages between 0.1 and 14 Gyr with a spacing of 0.1 Gyr between 0.1$-$2 Gyr, 0.5 Gyr between 2$-$6 Gyr and 1 Gyr between 6$-$14 Gyr. The fitting procedure results in a 2D solution giving the star formation rate~(SFR) for each age and metallicity. By projecting this solution onto the age axis we obtain the SFH, while a projection onto the metallicity axis gives the Chemical Evolution History~(CEH). 
\begin{figure*}[!ht]
\centering
\includegraphics[angle=0, width=0.97\textwidth]{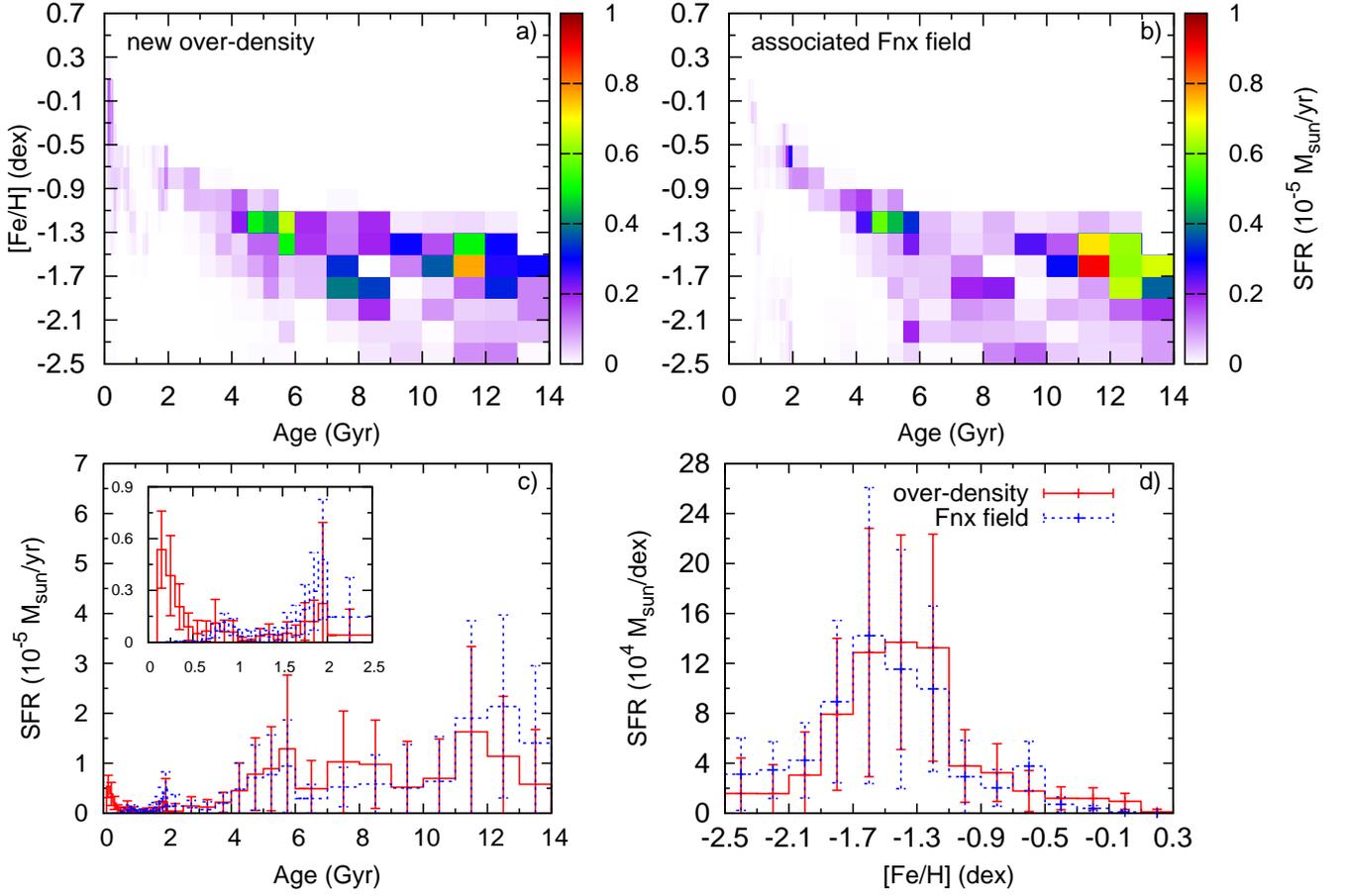}
\caption{The SFH of the newly discovered over-density, as obtained from the V,B-V CMD. See Figure~\ref{FnxSFHshell} for details. \label{FnxSFHshell2}} 
\end{figure*}
\\
Figure~\ref{FnxSFHshell} shows the SFH and CEH obtained for the known over-density discussed in Section~\ref{targetselection}. The SFH of the associated Fornax field contribution is also shown for comparison in Figure~\ref{FnxSFHshell}b. The SFH shows a peak at old~($\ge$10 Gyr) ages and a well-developed AMR at younger~($\le$6 Gyr) ages. The young populations in Figures~\ref{FnxSFHshell} show increased star formation at age$\approx$3 Gyr, [Fe/H]=$-$1.0 dex and a strong peak at age$\approx$1.5 Gyr with [Fe/H]$\approx$$-$0.6 dex. \\
The star formation episode at [Fe/H]=$-$1.0 dex, age$\approx$3 Gyr is consistent with the main metal-rich RGB of the Fornax field~(see Figure~\ref{shells}) and is recovered in both the over-density and associated field contribution. The younger peak at [Fe/H]=$-$0.6 dex, age 1.5 Gyr is only weakly reproduced in the Fornax field contribution. This population is clearly responsible for the stellar over-density observed in Figure~\ref{spatialBPhess}, showing parameters in excellent agreement with results from isochrone fitting of the observed CMD~\citep{Olszewski06}. Figure~\ref{FnxSFHshell}c further shows that the star formation rates at old~($\ge$10 Gyr) ages are consistent with those in the Fornax field at similar elliptical radii~(blue, dashed histogram) within the errorbars. This indicates that the over-density is formed purely by an excess of young, metal-rich stars without increased numbers of older stars, as suggested earlier by~\citet{Olszewski06}. \\
To calculate the total over-dense mass of stars in the young over-density, we  consider only stars younger than 2.5 Gyr and multiply the star formation rate in each population bin by the duration in age. This gives a total mass in young stars of 5.75$\pm$1.31$\times$10$^{4}$ M$_{\odot}$ for the over-density and 1.27$\pm$0.65$\times$10$^{4}$ M$_{\odot}$ for the associated Fornax field contribution. Subtracting the field star contribution gives a total over-dense mass of 4.47$\pm$1.46$\times$10$^{4}$ M$_{\odot}$ formed within a mass range of 0.1$-$120~M$_{\odot}$, assuming a Kroupa IMF. Compared to the stellar mass at all ages~(9.92$\pm$2.08$\times$10$^{4}$ M$_{\odot}$), the young populations account for $\approx$45 percent of the total stellar mass formed within the region selected in Section~\ref{targetselection}. \\
Figure~\ref{FnxSFHshell2} shows the full SFH solution of the newly discovered young over-density. The young over-density displays populations with a slowly increasing metallicity for younger ages. The AMR shows continuous evolution up to [Fe/H]=$-$0.9 dex, age$\approx$3 Gyr, with later star formation episodes at [Fe/H]=$-$0.7 dex, age$\approx$2 Gyr and [Fe/H]=0.0 dex, age$\approx$200 Myr. The peak at [Fe/H]=$-$0.7 dex displays similar parameters as the main peak in the known over-density, suggesting a possible connection between both features. However, the peak in the young over-density is shifted to slightly older ages as observed in Figure~\ref{FnxSFHshell}. The SFH solution of the Fornax field~(top, right panel in Figure~\ref{FnxSFHshell2}) shows a similar AMR as the young over-density, except for the peak at the youngest age. This is also visible in the bottom panels of Figure~\ref{FnxSFHshell2}, which show that the SFH and CEH of stars with age$\ge$1 Gyr are reproduced by the field star contribution, within the errorbars. The SFH shows that the population responsible for the over-density in Figure~\ref{spatialBPhess} is clearly extremely young, with ages between 100$-$300 Myr and metallicities up to [Fe/H]$\approx$0.0 dex. \\
For the total over-dense mass in stars we consider only stars with ages $\le$1 Gyr and compute a total mass in young stars of 1.56$\pm$0.41$\times$10$^{4}$ M$_{\odot}$ within the over-density, and 0.30$\pm$0.14$\times$10$^{4}$ M$_{\odot}$ within the Fornax field. Relative to the Fornax field contribution, the young over-density formed a total mass of 1.26$\pm$0.43$\times$10$^{4}$ M$_{\odot}$ in stars during the last Gyr, within an area of $\approx$6 arcmin$^{2}$. This corresponds to $\approx$10 percent of the total stellar mass of 1.32$\pm$0.37$\times$10$^{5}$ M$_{\odot}$ in Figure~\ref{FnxSFHshell2}a.

\section{Comparison to the Fornax centre}
\label{centrecomparison}
\begin{figure}[!ht]
\centering
\includegraphics[angle=0, width=0.49\textwidth]{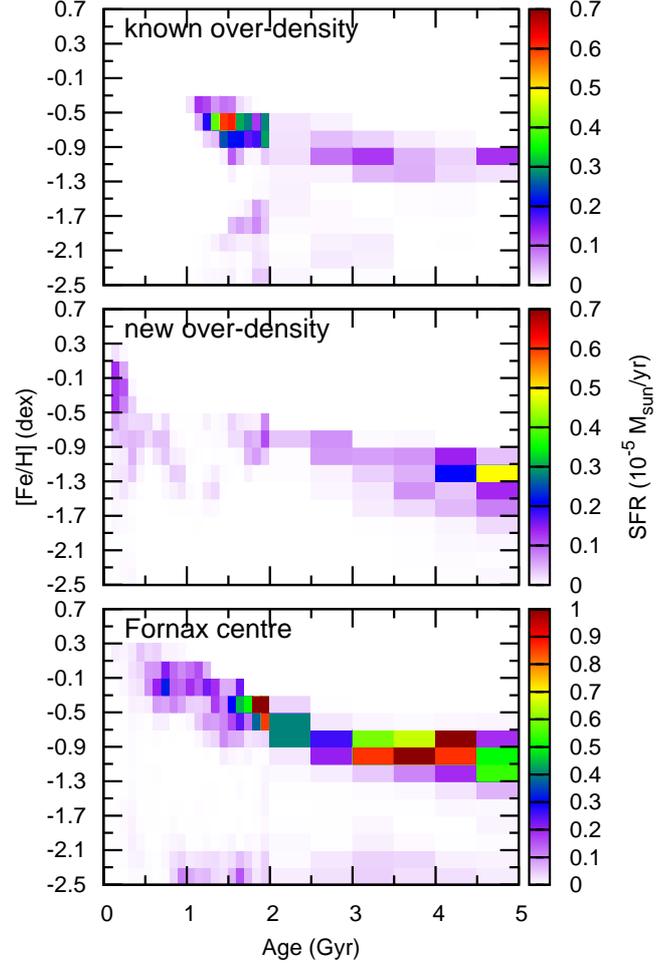}
\caption{Comparison between the young star formation~(age$\le$5 Gyr) in the known over-density, new over-density and the centre of the Fornax dSph~(r$_{ell}$$\le$0.194 degrees). The SFH solution obtained for the Fornax centre has been re-scaled to the same spatial area as the known over-density. \label{Fnxcentrecomparison}} 
\end{figure}

\begin{figure}[!ht]
\centering
\includegraphics[angle=0, width=0.49\textwidth]{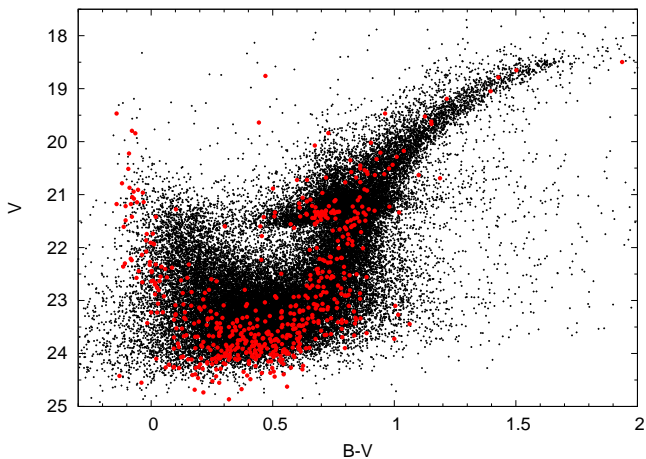}
\caption{Comparison between the CMD of the centre of the Fornax dSph for r$_{ell}$$\le$0.194 degrees~(black crosses) and the CMD of the new young stellar over-density~(red filled circles). \label{FnxCMDcomparison}} 
\end{figure}

The youngest populations of the Fornax field have sofar been found in the centre of the galaxy~\citep[e.g.,][]{Gallart052, Coleman08, deBoer2012B}. If the observed stellar over-densities were created by the accretion of a gas-rich companion~\citep{Coleman04}, the in-fall event likely triggered the young star formation in the central regions. This would lead to the presence of similar stellar populations in the over-densities and the Fornax centre. Furthermore, by comparing the properties of both components, we can determine if the stars in the over-densities match the enrichment history of the Fornax dSph, indicating that they were formed from Fornax gas. \\
To determine if the populations of the stellar over-densities match those of the youngest Fornax field populations, we compare them to the stellar population properties of the stars within the innermost annulus considered in~\citet{deBoer2012B}. We re-determine the SFH using the same method and parameter spacing as adopted for the over-densities~(see Section~\ref{FnxSFH}), and scale the solution to the same spatial area as the known over-density to allow a quantitative comparison. Figure~\ref{Fnxcentrecomparison} shows the resulting SFH of the Fornax centre in comparison to the stellar over-densities, for ages younger than 5 Gyr. 
\\
The centre of Fornax shows a narrow AMR at older ages, from [Fe/H]=$-$1.3, age=5 Gyr down to [Fe/H]=$-$0.7, age$\approx$2 Gyr. This is followed by a change in slope in the AMR and the presence of more metal-rich populations up to [Fe/H]=0.0 dex at age$\approx$500 Myr. The tight AMR between ages 2$-$5 Gyr is also reproduced in both stellar over-densities, and likely belongs to the Fornax field~(see also Figures~\ref{FnxSFHshell} and~\ref{FnxSFHshell2}). In the centre of Fornax, the star formation rates go down along the narrow AMR with decreasing age, followed by an increase at age$\approx$2 Gyr, [Fe/H]=$-$0.5 dex. This coincides with the main star formation peak in the known over-density, hinting at a link between the two components. In the case of an in-fall event, the gas that formed the over-density may also have triggered new star formation in the Fornax centre around age=1.5 Gyr. \\
The youngest stars in the Fornax field AMR display age$\approx$500 Myr and [Fe/H]$\approx$+0.1 dex, marking the termination of star formation in the centre. Comparison to the middle panel of Figure~\ref{Fnxcentrecomparison} shows that the very young populations in the new over-density are not reproduced in the Fornax centre or the known over-density. Comparison between the CMD of the Fornax centre and the young over-density~(Figure~\ref{FnxCMDcomparison}) shows that stars consistent with the young MS could be present in the Fornax centre, but only at very low levels. Therefore, the new over-density formed after the last substantial episode of star formation in the Fornax centre, with a metallicity comparable to the youngest Fornax field stars. 

\section{Discussions and conclusions}
\label{conclusions}
We have conducted a new study of young stellar over-densities in the Fornax dSph galaxy, using wide-field, deep photometric data presented in~\citet{deBoer2012B}. By making a colour selection to isolate young stars, we have discovered a new stellar over-density in Fornax, located at a distance of 0.3 degrees~(0.7 kpc) from the centre~(see Figure~\ref{spatialBPhess}). Using deep V,B$-$V CMDs, we have analysed the stellar populations of the known over-density and the newly discovered over-density. \\
The previously discovered over-dense feature displays a CMD with a young MS feature that is only weakly reproduced in the associated Fornax field star contribution. The SFH~(see Figure~\ref{FnxSFHshell}) shows a well-developed AMR at young ages and a main over-dense population with metallicity~[Fe/H]$\approx$$-$0.6 dex and age$\approx$1.5 Gyr. These values are consistent with the young MS feature in the CMD as well as previous studies of the over-density~\citep{Olszewski06}. Comparison between the centre of Fornax and the over-density~(see Figure~\ref{Fnxcentrecomparison}) shows that the main population is consistent with the Fornax AMR, although located at a relatively large distance from the centre~(1.3 r$_{core}$). The high metallicity of the stars makes it unlikely that the over-density is the result of an in-falling primordial gas cloud. Furthermore, since there are no signs of increased numbers of old and intermediate stars, it is unlikely that it was formed by the accretion of a small gas-rich galaxy, unless it contained very little star formation at the time of in-fall. \\
The properties of the known over-density are consistent with having been formed by the re-accretion of previously expelled gas. Such a scenario has been proposed by~\citet{Yozin12} to explain the multiple episodes of star formation in the Carina and Fornax dSphs, and would explain why the gas is enriched to similar values as the Fornax field. Furthermore, an in-fall event is the most likely process to explain the presence of a young outer over-dense feature outside of the tidal radius~\citep{Coleman05}. If both the inner and outer over-densities were the result of the same process, the AMR of both features will be consistent. This can be determined by conducting a deep CMD study of the outer structure discovered by~\citet{Coleman05}. \\
The new over-density is also formed of young stars. The CMD~(Figure~\ref{shells}) shows a very young MS that is not reproduced in the field stars. The SFH in Figure~\ref{FnxSFHshell2} shows a clear peak at ages $\le$0.3 Gyr with metallicities close to solar~([Fe/H]=0.0 dex), confirming that this over-density constitutes some of the youngest, most metal-rich star formation in Fornax. The SFH of the new over-density is peaked in the youngest age bin considered in this study~(100 Myr), implying that the stars could be even younger than the age determined here. However, the excellent fit of the 100$-$200 Myr isochrone to the young MS~(see Figure~\ref{shells}) gives confidence to the age recovered in the SFH. However, the use of deeper CMDs and spectroscopic metallicity measurements would allow a more accurate determination of the age of this young over-density. \\
It is unclear what triggered this young star formation episode, especially given its location outside the centre of Fornax. If both previously detected over-dense features are linked to a recent in-fall event, it seems possible the new over-density may also be linked to an accretion event. The total mass in stars in the young populations is relatively small~($\approx$10$^{4}$ M$_{\odot}$) suggesting it could be formed from Fornax gas or the accretion of a small gas cloud. However, since the metallicity of stars in the over-density is comparable to the youngest Fornax stars, it is more likely that the feature formed from leftover gas expelled during the last episode of star formation in the Fornax centre.\\
The SFH of both stellar over-densities show the presence of stellar populations with ages 1.5$-$2.0 Gyr. Furthermore, evidence of this population is also found in the associated Fornax field and the central region of Fornax. Therefore, this population could be more widespread than just the over-density where it was first found. If this feature was indeed formed by an in-fall event it is possible that stars with these properties were spread throughout the Fornax dSph during the accretion process. However, kinematic information is needed to unambiguously determine the origin and connection of both over-densities and their association to the young stellar populations of Fornax.

\section{Acknowledgements}
\label{acknowledgements}
The authors thank ISSI (Bern) for support of the team ``Defining the full life-cycle of dwarf galaxy evolution: the Local Universe as a template". T.d.B and E.T. gratefully acknowledge the Netherlands Foundation for Scientific Research (NWO) for financial support through a VICI grant. E.O. is partially supported by NSF grant AST0807498. The authors would like to thank the anonymous referee for his/her comments, that helped to improve the paper.

\bibliographystyle{aa}
\bibliography{Bibliography}

\end{document}